\documentclass[11pt,twoside,onecolumn]{article}
\usepackage[]{latexsym}
\usepackage{epsfig}
\usepackage{amsmath,amssymb}
\RequirePackage[dvipsnames,usenames]{color}
\newcommand{\be}{\begin{eqnarray}}
\newcommand{\ee}{\end{eqnarray}}

\title{\bf The Minimal Geometric Deformation Approach:
\\
a brief introduction}
\author{J Ovalle$^{ab}$\thanks{jovalle@usb.ve}
$\,$ 
R Casadio$^{cd}$\thanks{casadio@bo.infn.it}
$\,$ 
A Sotomayor$^{e}$\thanks{adrian.sotomayor@uantof.cl}
\\
\null
\\
$^a${\em Departamento de F\'{\i}sica, Universidad Sim\'on Bol\'ivar,}
\\
{\em AP 89000, Caracas 1080A, Venezuela}
\\
$^b${\em The Institute for Fundamental Study, Naresuan University}
\\
{\em Phitsanulok 65000, Thailand}
\\
$^c${\em Dipartimento di Fisica e Astronomia, Alma Mater Universit\`a di Bologna}
\\
{\em via Irnerio~46, 40126 Bologna, Italy}
\\
$^d${\em Istituto Nazionale di Fisica Nucleare, Sezione di Bologna, I.S.~FLAG}
\\
{\em  viale Berti~Pichat~6/2, 40127 Bologna, Italy}
\\
$^e${\em Departamento de Matem\'aticas, Universidad de Antofagasta}
\\
{\em  Antofagasta, Chile}
}
\begin{document}
\maketitle
\begin{abstract}

We review the basic elements of the Minimal Geometric Deformation approach in details.
This method has been successfully used to generate brane-world configurations from 
general relativistic perfect fluid solutions.
\end{abstract}
%
%
%
%
%
%
%
\section{Introduction}
General Relativity (GR) represents one of the pillars of modern Physics.
The predictions made by this theory range from the perihelion shift of Mercury, the deflexion of light
and gravitational lensing, the gravitational redshift and time delay, and the existence of black holes.
The observation of these effects, as well as the recent detection of the gravitational waves
GW150914~\cite{ligo1} and GW151226~\cite{ligo2}, have given GR the status of the benchmark
theory of the gravitational interaction (for an excellent review, see Ref.~\cite{cmw} and references therein).
Why do we want to find new gravitational theories beyond GR then?
The reason has to do with some fundamental questions associated with the gravitational interaction
which GR does not seem to be able to answer satisfactorily.
One is the problem of dark matter and dark energy, which require introducing some unknown
matter-energy to reconcile GR predictions with the observations of galactic rotation curves and
accelerated expansion of the universe, respectively.
Then, there is the difficulty of reconciling GR with the Standard Model of particle physics, 
or equivalently, the failure to quantise GR by the same successful scheme used with the other
fundamental interactions.
Such issues have motivated the search of a new gravitational theory beyond GR that could help
to explain part of the problems mentioned above.
Indeed, there is already a long list of alternative theories, like $f(R)$ and higher curvature theories,
Galileon theories, scalar-tensor theories, (new and topological) massive gravity,
Chern-Simons theories, higher spin gravity theories, Horava-Lifshitz gravity, extra-dimensional theories,
torsion theories, Horndeski's theory, etc (See for instance Refs.~\cite{Ber}--\cite{Bellorin}).
Nonetheless, quantum gravity is still an open problem, and dark matter and dark energy remain a mystery
so far.
\par
The MGD was originally proposed~\cite{jo1} in the context of the the Randall-Sundrum
brane-world~\cite{lisa1,lisa2} and extended to investigate new black hole
solutions~\cite{MGDextended1,MGDextended2}.
While the brane-world is still an attractive scenario, since it explains the hierarchy of fundamental 
interactions in a simple way, to find interior solutions for self-gravitating systems is a difficult task,
mainly due to the existence of non-linear terms in the matter fields.
In addition, the effective four-dimensional Einstein equations are not a closed system,
due to the extra-dimensional effects resulting in terms undetermined by the four-dimensional
equations.
Despite these complications, the MGD has proven to be useful, among other things, to derive
exact and physically acceptable solutions for spherically symmetric and non-uniform stellar
distributions~\cite{jo2,jo5} as well; to express the tidal charge in the metric found in
Ref.~\cite{dmpr} in terms of the usual Arnowitt-Deser-Misner (ADM) mass~\cite{jo6};
to study microscopic black holes~\cite{jo7}; to clarify the role of exterior Weyl stresses acting on
compact stellar distributions~\cite{jo8,jo9}; to extend the concept of variable tension
introduced in Refs.~\cite{gergely2009} by analysing the shape of the black string in the
extra dimension~\cite{jo10}; to prove, contrary to previous claims, the consistency
of a Schwarzschild exterior~\cite{jo11} for a spherically symmetric self-gravitating
system made of regular matter in the brane-world; to derive bounds on extra-dimensional
parameters~\cite{jo12} from the observational results of the classical tests of GR in the
Solar system;
to investigate the gravitational lensing phenomena beyond GR~\cite{roldaoGL}; to determine the critical stability region for Bose-Einstein condensates
in gravitational systems~\cite{rrplb}; to study Dark $SU(N)$ glueball stars on fluid branes~\cite{rolsun} as well as the correspondence between sound waves in a de Laval propelling nozzle and quasinormal modes emitted by brane-world black holes~\cite{rol}. 
\par
This brief review is organised as follows:
the simplest ways to modified gravity are presented in Section~\ref{s2},
emphasising some problems that arise when the GR limit is considered;
in Section~\ref{s3}, we recall the Einstein field equations on the brane for a spherically symmetric
and static distribution of density $\rho$ and pressure $p$;
in Section~\ref{s4}, the GR limit is discussed and the basic elements of the MGD are
presented in section~\ref{s5};
in Section~\ref{s6}, we review the matching conditions between the interior and exterior space-time of
self-gravitating systems within the MGD, and a recipe with the basic steps to implement the MGD is
described in Section~\ref{s7};
finally, some conclusions are presented in the last section.
\section{GR simplest extensions and their GR limit}
\label{s2}
\setcounter{equation}{0}
This Section is devoted to describe in a qualitative way the so-called GR-limit problem, which arises when an extension to GR is considered. An explicit and quantitative description of this problem, as well as an explicit solution, is developed throughout the rest of the review.
\par
One cannot try and change GR without considering the well-established and very useful
Lovelock's theorem~\cite{lovelock}, which severely restricts any possible ways of modifying
GR in four dimensions.
We will now show the simplest generic way.
\par
Any extension to GR will eventually produce new terms in the effective four-dimensional Einstein equations.
These ``corrections''  are usually handled as part of an effective energy-momentum tensor and appear
in such a way that they should vanish or be negligible in an appropriate limit.
For instance, they must vanish (or be negligible) at solar system scales, where GR has been successfully tested so far.\footnote{Of course any deviation from GR/Newton theory at i) very short distances or ii) beyond the Solar System scale is welcome as long as it could deal with the quantum  problem or dark matter problem}
This limit represents not only a critical point for a consistent extension of GR, but also a non-trivial problem
that must be treated carefully.
\par
\begin{figure}[t]
\center
\includegraphics[scale=0.4]{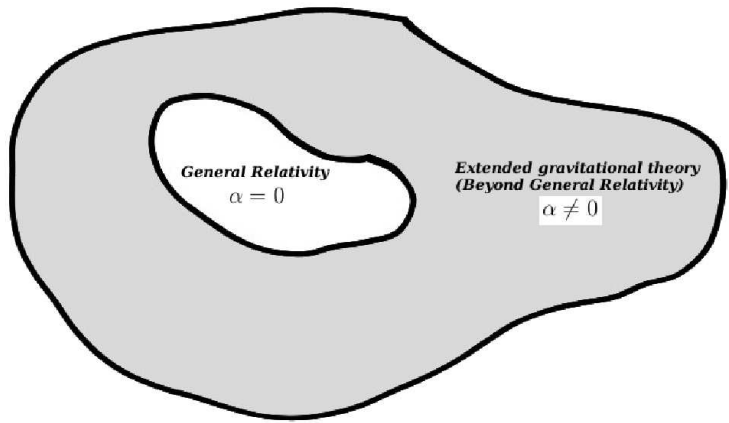}
\\
\centering\caption{The new gravitational sector outside GR is parametrised by $\alpha$,
so that GR represents a ``sub-space" in the extended theory of gravity.
When the free parameter $\alpha$ is turned off, we should automatically recover the domain of GR.}
\label{fig:extended}      
\end{figure}

The simplest way to extend GR is by considering a modified Einstein-Hilbert action,
\begin{equation}
\label{corr1}
S
=
\int\left[\frac{R}{2\,k^2}+{\cal L}\right]
\sqrt{-g}\,d^4x
+\alpha\,({\rm correction})
\ ,
\end{equation} 
where $\alpha$ is a free parameter associated with the new gravitational sector not described by GR, as is  schematically shown in Fig \ref{fig:extended}.
The explicit form corresponding to the generic correction shown in Eq.~(\ref{corr1}) should be, of course,
a well justified and physically motivated expression.
At this stage the GR limit, obtained by setting $\alpha = 0$, is just a trivial issue, so everything looks
consistent.
Indeed, we may go further and calculate the equations of motion from setting the variation
$\delta S = 0$ corresponding to this new theory,
\begin{equation}
\label{corr2}
R_{\mu\nu}-\frac{1}{2}\,R\, g_{\mu\nu}
=
k^2\,T_{\mu\nu}
+\alpha\,({\rm new\ terms})_{\mu\nu}
\ .
\end{equation}
The new terms in Eq.~(\ref{corr2}) may be viewed as part of an effective energy-momentum tensor,
whose explicit form may contain new fields, like scalar, vector and tensor fields,
all of them coming from the new gravitational sector not described by Einstein's theory.
At this stage the GR limit, again, is a trivial issue, since $\alpha = 0$ leads to the standard Einstein's equations 
$G_{\mu\nu}=k^2\,T_{\mu\nu}$.
\par
All the above seems to tell us that the consistency problem, namely the GR limit, is trivial.
However, when the system of equations given by the expression~(\ref{corr2}) is solved,
the result may show a complete different story.
In general, and this is very common, the solution eventually found cannot reproduce the GR limit by
simply setting $\alpha = 0$.
The cause of this problem is the non-linearity of Eq.~(\ref{corr2}), and should not be a surprise.
To clarify this point, let us consider a spherically symmetric perfect fluid, 
for which GR uniquely determines the metric component
\begin{equation}
\label{g11-1}
g^{-1}_{rr} = 1 - \frac{2\, m(r)}{r}
\ ,
\end{equation}
where $m$ is the mass function of the self-gravitating system.
Now, let us consider the same perfect fluid in the ``new'' gravitational theory~(\ref{corr1}).
When Eq.~(\ref{corr2}) is solved, we obtain an expression which generically may be written as
\begin{equation}
\label{g11def}
g^{-1}_{rr} = 1 - \frac{2\, m(r)}{r} + ({\rm geometric\ deformation})
\ ,
\end{equation}
where by {\it geometric deformation\/} one should understand the deformation of
the metric component~(\ref{g11-1}) due to the generic extension~(\ref{corr1}) of GR.
(``deformation'' hence means a deviation from the GR solution).
It is now very important to note that the deformation~(\ref{g11def}) always produces
{\it anysotropic consequences\/} on the perfect fluid, namely, the radial and tangential pressures are no longer the same and in consequence the self-gravitating system
will not be described as a perfect fluid anymore.
Indeed, and this is a critical point in our analysis, the anisotropy produced by the 
geometric deformation always takes the form (See further Eqs.~(\ref{ppf})-(\ref{ppf3}) to see an explicit calculation)
\begin{equation}
\label{any}
{\cal P} = A + \alpha\,B
\ .
\end{equation}
This expression is very significant, since it shows that the GR limit cannot be  {\em a posteriori\/}
recovered by setting $\alpha=0$, since the ``sector'' denoted by $A$ in the anisotropy~(\ref{any}) 
does not depend on $\alpha$.
Consequently, the perfect fluid GR solution ($A=0$) is not trivially contained in this extension,
and one might say that we have an extension to GR which does not contain GR.
This is of course a contradiction, or more properly a consistency problem, whose
source can be precisely traced back to the {\it geometric deformation} shown in Eq.~(\ref{g11def}).
The latter always takes the form (See Eq.~(\ref{fsolution}) for an explicit expression)
\begin{equation}
\label{def}
({\rm geometric\ deformation}) = X + \alpha\,Y
\ ,
\end{equation}
which contains a ``sector'' $X$ that does not depend on $\alpha$.
This is again obviously inconsistent, since the deformation undergone by GR must 
depend smoothly on $\alpha$ and vanish with it.
At the level of solutions, the source of this problem is the high non-linearity of the effective
Einstein equations~(\ref{corr2}), which we want to emphasise has nothing to do with 
any specific extension of GR.
Indeed, it is a characteristic of any high non-linear systems.
\par
\begin{figure}[t]
\center
\includegraphics[scale=0.4]{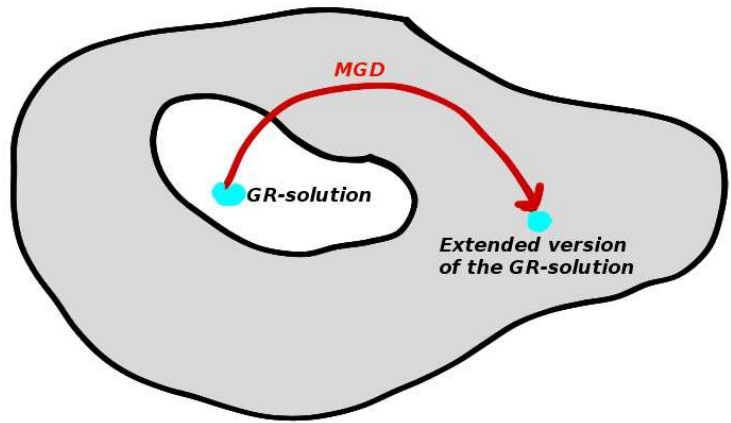}
\\
\centering\caption{When a GR solution is forced to be a solution in the new gravitational sector by the MGD,
the $\alpha$-independent terms in the extended solution are eliminated, and the GR limit is recovered.}
\label{fig:MGD}      
\end{figure}

A method that solves the non-trivial issue of consistency with GR described above is the so-called
{\it Minimal Geometric Deformation} MGD approach~\cite{jo1}. 
The idea is to keep under control the anisotropic consequences on GR appearing in the extended theory,
in such a way that the $\alpha$-independent sector in the geometric deformation shown as $X$
in Eq.~(\ref{def}) always vanishes.
Correspondingly, the $\alpha$-independent sector of the anisotropy $A$ in Eq. (\ref{any}) will
also vanish.
This will ensure a consistent extension that recovers GR in the limit $\alpha\to 0$.
In this approach, the generic expression $Y$ in Eq.~(\ref{def}) represents the
{\it minimal geometric deformation} undergone by the radial metric component,
the generic expression $B$ in Eq.~(\ref{any}) being the {\it minimal anysotropic consequence}
undergone by GR due to correction terms in the modified Einstein-Hilbert action~(\ref{corr1}). 
The next key point is how we can make sure $X = 0$ in Eq.~(\ref{def}) in order to obtain a consistent
extension to GR. 
This is accomplished when a GR solution is forced to remian a solution in the extended theory.
Roughly speaking, we need to introduce the GR solution into the new theory, as far as possible, as suggested in Fig~\ref{fig:MGD}. This provides the foundation for the MGD approach.
We want to emphasise that the GR solution used to set $X = 0$ in Eq.~(\ref{def}) will eventually
be modified by using, for instance, the matching conditions at the surface of a self-gravitating system. 
One will therefore obtain physical variables that depend on the free parameter of the theory,
here generically named $\alpha$.
This free parameter could be, for instance, the one that measures deviation from GR in $f(R)$
theories, the brane tension in the brane-world, and so. 
\section{Extra-dimensional gravity: the brane-world}
\label{s3}
\setcounter{equation}{0}
In the generalised RS brane-world scenario, gravity lives in five dimensions and
affects the gravitational dynamics in the (3+1)-dimensional universe accessible to all
other physical fields, the so-called brane.
The 5-dimensional Einstein equations projected on the brane give rise to the modified
4-dimensional Einstein equations~\cite{maartRev2004x,maartRev2010x,smsx}~\footnote{We
use units with $G$ the 4-dimensional Newton constant, $k^{2}=8\,\pi \,G$, and $\Lambda$
the 4-dimensional cosmological constant.}
\begin{equation}
G_{\mu \nu }
=
-k^{2}\,T_{\mu \nu }^{\mathrm{eff}}-\Lambda \,g_{\mu \nu }
\ ,
\label{4Dein}
\end{equation}
where $G_{\mu\nu}$ is the 4-dimensional Einstein tensor.
The effective energy-momentum tensor is given by
\begin{equation}
T_{\mu \nu }^{\mathrm{eff}}
=
T_{\mu \nu }
+\frac{6}{\sigma }\,S_{\mu \nu }+\frac{1}{8\,\pi}\,\mathcal{E}_{\mu \nu }
+\frac{4}{\sigma }\,\mathcal{F}_{\mu \nu }
\ ,
\label{tot}
\end{equation}
where $\sigma $ is the brane tension (which plays the role of the parameter $\alpha$
of the previous Section) and 
\begin{equation}
T_{\mu \nu }=(\rho +p)\,u_{\mu }\,u_{\nu }-p\,g_{\mu \nu }
\label{perfect}
\end{equation}
is the 4-dimensional energy-momentum tensor of brane matter described by
a perfect fluid with 4-velocity field $u^\mu$, density $\rho$ and isotropic pressure $p$.
The extra term 
\begin{equation}
S_{\mu \nu }
=
\frac{T}{12}\,T_{\mu \nu}
-\frac{1}{4}\,T_{\mu \alpha }\,T_{\ \nu}^{\alpha }
+\frac{g_{\mu \nu }}{24}\left( 3\,T_{\alpha \beta}\,T^{\alpha \beta }-T^{2}\right)
\end{equation}
represents a local high-energy correction quadratic in $T_{\mu\nu}$ (with $T=T_{\alpha }^{\ \alpha }$),
whereas
\begin{equation}
k^{2}\,\mathcal{E}_{\mu \nu }
=
\frac{6}{\sigma }\left[
\mathcal{U}
\left(
u_{\mu }\,u_{\nu }
+\frac{1}{3}\,h_{\mu \nu }\right)
+\mathcal{P}_{\mu \nu }
+\mathcal{Q}_{(\mu }\,u_{\nu )}\right]
\end{equation}
contains the Kaluza-Klein corrections and acts as a non-local source arising from the
5-dimensional Weyl curvature.
Here $\mathcal{U}$ is the bulk Weyl scalar, $\mathcal{P}_{\mu \nu }$ the Weyl stress tensor and 
$\mathcal{Q}_{\mu }$ the Weyl energy flux, and $h_{\mu\nu}=g_{\mu\nu}-u_\mu u_\nu$ denotes
the projection tensor orthogonal to the fluid lines.
Finally, the extra term $\mathcal{F}_{\mu \nu }$ contains contributions from all non-standard
model fields possibly living in the bulk, but it does not include the 5-dimensional
cosmological constant $\Lambda_5$, which is fine-tuned to $\sigma$ in order to generate
a small 4-dimensional cosmological constant
\begin{equation}
\Lambda
=
\frac{\kappa_5^{2}}{2}\left(\Lambda_{5}+\frac{1}{6}\,\kappa_5^{2}\,\sigma^{2}\right)
\simeq
0
\ ,
\end{equation}
where the 5-dimensional gravitational coupling 
\begin{equation}
\kappa^4_5
=
6\,\frac{k^2}{\sigma}
\ .
\end{equation}
In particular, we shall only allow for a cosmological constant in the bulk, hence
\begin{equation}
{\cal F}_{\mu\nu}=0
\ ,
\end{equation}
which implies the conservation equation
\begin{equation}
\nabla_\nu\,T^{\mu\nu}=0
\ ,
\label{dT0}
\end{equation}
and there will be no exchange of energy between the bulk and the brane. 
\par
In this review, we are mostly interested in spherically symmetric and static configurations,
for which the Weyl energy flux
\begin{equation}
Q_\mu =0
\end{equation}
and the Weyl stress can be written as
\begin{equation}
{\cal P}_{\mu\nu}
={\cal P}\left(r_\mu\, r_\nu+\frac{1}{3}\,h_{\mu\nu}\right)
\ ,
\end{equation}
where $r_\mu$ is a unit radial vector.
In Schwarzschild-like coordinates, the spherically symmetric metric reads 
\begin{equation}
ds^{2}
=
e^{\nu (r)}\,dt^{2}-e^{\lambda (r)}\,dr^{2}
-r^{2}\left( d\theta^{2}+\sin ^{2}\theta \,d\phi ^{2}\right)
\ ,
\label{metric}
\end{equation}
where $\nu =\nu (r)$ and $\lambda =\lambda (r)$ are functions of the areal
radius $r$ only, ranging from $r=0$ (the star's centre) to some $r=R$ (the
star's surface), and the fluid 4-velocity field is given by
$u^{\mu }=e^{-\nu /2}\,\delta _{0}^{\mu }$ for $0\le r\le R$.
\par
The metric~(\ref{metric}) must satisfy the effective Einstein equations~(\ref{4Dein}),
which, for $\Lambda=0$, explicitly read~\cite{germ,matt,fran}
\begin{eqnarray}
\label{ec1}
&&
k^2
\left[ \rho
+\strut\displaystyle\frac{1}{\sigma}\left(\frac{\rho^2}{2}+\frac{6}{k^4}\,\cal{U}\right)
\right]
=
\strut\displaystyle\frac 1{r^2}
-e^{-\lambda }\left( \frac1{r^2}-\frac{\lambda'}r\right)
\\
&&
\label{ec2}
k^2
\strut\displaystyle
\left[p+\frac{1}{\sigma}\left(\frac{\rho^2}{2}+\rho\, p
+\frac{2}{k^4}\,\cal{U}\right)
+\frac{4}{k^4}\frac{\cal{P}}{\sigma}\right]
=
-\frac 1{r^2}+e^{-\lambda }\left( \frac 1{r^2}+\frac{\nu'}r\right)
\\
&&
\label{ec3}
k^2
\strut\displaystyle\left[p
+\frac{1}{\sigma}\left(\frac{\rho^2}{2}+\rho\, p
+\frac{2}{k^4}\cal{U}\right)
-\frac{2}{k^4}\frac{\cal{P}}{\sigma}\right]
=
\frac 14e^{-\lambda }\left[ 2\,\nu''+\nu'^2-\lambda'\,\nu'
+2\,\frac{\nu'-\lambda'}r\right]
\ .
\end{eqnarray}
Moreover, the conservation Eq.~(\ref{dT0}) yields
\begin{equation}
\label{con1}
p'=-\strut\displaystyle\frac{\nu'}{2}(\rho+p)
\ ,
\end{equation}
where $f'\equiv \partial_r f$.
We then note the 4-dimensional GR equations are formally
recovered for $\sigma^{-1}\to 0$, and the conservation equation~(\ref{con1})
then becomes a linear combination of Eqs.~(\ref{ec1})-(\ref{ec3}).
\par
By simple inspection of the field equations~(\ref{ec1})-(\ref{ec3}), we
can identify an effective density 
\begin{equation}
\tilde{\rho}
=
\rho
+\frac{1}{\sigma }
\left( \frac{\rho ^{2}}{2}+\frac{6}{k^{4}}\,\mathcal{U}\right)
\ ,
\label{efecden}
\end{equation}
an effective radial pressure
\begin{equation}
\tilde{p}_{r}
=
p
+\frac{1}{\sigma }\left( \frac{\rho ^{2}}{2}+\rho \,p
+
\frac{2}{k^{4}}\,\mathcal{U}\right)
+\frac{4}{k^4\,\sigma }\,\mathcal{P}
\ ,
\label{efecprera}
\end{equation}
and an effective tangential pressure
\begin{equation}
\tilde{p}_{t}
=
p+\frac{1}{\sigma }\left( \frac{\rho ^{2}}{2}+\rho \,p
+\frac{2}{k^{4}}\,\mathcal{U}\right)
-\frac{2}{k^{4}\,\sigma}\,\mathcal{P}
\ .
\label{efecpretan}
\end{equation}
This clearly illustrates that extra-dimensional effects generate an anisotropy 
\begin{equation}
\Pi
\equiv
\tilde{p}_{r}-\tilde{p}_{t}
=
\frac{6}{k^{4}\,\sigma}\,\mathcal{P}
\ ,
\end{equation}
inside the stellar distribution.
A GR isotropic stellar distribution (perfect fluid) therefore becomes an
anysotropic stellar system on the brane.
\par
Eqs.~(\ref{ec1})-(\ref{con1}) contain six unknown functions, namely:
two physical variables, the density $\rho(r)$ and pressure $p(r)$;
two geometric functions, the temporal metric function $\nu(r)$ and the radial function $\lambda(r)$;
and two extra-dimensional fields, the Weyl scalar function ${\cal U}$ and the anisotropy ${\cal P}$.
These equations therefore form an indefinite system on the brane, an open problem to solve which
one needs more information about the bulk geometry and a better understanding of how
our 4-dimensional spacetime is embedded in the bulk~\cite{cmazza,darocha2012}.
Since the source of this problem is directly related to the projection ${\cal E}_{\mu\nu}$ of the bulk
Weyl tensor on the brane, the first logical step to overcome this issue would be to impose the constraint
${\cal E}_{\mu\nu}=0$ on the brane.
However, it was shown in Ref.~\cite{koyama05} that this condition is incompatible with the Bianchi
identity on the brane, and a different and less radical restriction must thus be implemented.
Another option that has led to some success consists in discarding only the anisotropic stress
associated to ${\cal E}_{\mu\nu}$, that is, setting ${\cal P}_{\mu\nu}=0$.
This constraint, which is useful to overcome the non-closure problem~\cite{shtanov07}, 
is nonetheless still too strong, since some anisotropic effects on the brane are generically expected
as a consequence of the ``deformation'' induced on the 4-dimensional geometry by 5-dimensional
gravity~\cite{jo1}.
\section{The GR limit in the brane-world}
\setcounter{equation}{0}
\label{s4}
Despite the non-closure problem that plagues the effective 4-dimensional Einstein equations,
we shall see that it is possible to determine a brane-world version of every known GR
perfect fluid solution.
In order to do so, the first step is to rewrite the equations~(\ref{ec1})-(\ref{ec3}) in a suitable
way.
First of all, 
by combining Eqs.~(\ref{ec2}) and~(\ref{ec3}), we obtain the Weyl anysotropy 
\begin{equation}
\label{pp}
\frac{6\,{\cal P}}{k^2\,\sigma}
=
G_{\ 1}^{1}-G_{\ 2}^2
\end{equation}
and Weyl scalar
\begin{equation}
\frac{6\,{\cal U}}{k^4\,\sigma}
=
-\frac{3}{\sigma}\left(\frac{\rho^2}{2}+\rho\,p\right)
+\frac{1}{k^2}\left(2\,G_{\ 2}^2+G_{\ 1}^1\right)-3\,p
\ ,
\label{uu}
\end{equation}
where
\begin{equation}
\label{g11}
G_{\ 1}^1
=
-\frac 1{r^2}+e^{-\lambda }\left( \frac 1{r^2}+\frac{\nu'}r\right)\ ,
\end{equation}
and
\begin{equation}
\label{g22}
G_{\ 2}^2
=
\frac 14\,e^{-\lambda }\left( 2\,\nu''+\nu'^2-\lambda'\,\nu'+2\, \frac{\nu'-\lambda'}{r}
\right)
\ .
\end{equation}
Eqs.~(\ref{pp})-(\ref{g22}) are equivalent to Eqs.~(\ref{ec1})-(\ref{con1}) and
we still have an open system of equations for the three unknown functions
$\{p, \rho, \nu\}$ satisfying the conservation equation~(\ref{con1}).
Next, we can proceed by plugging Eq.~(\ref{uu}) into Eq.~(\ref{ec1}),
which leads to a first order linear differential equation for the metric function $\lambda$, 
\begin{eqnarray}
\label{edlrw}
e^{-\lambda}
\left(r\,\lambda'-1\right)
+1
-r^2\,k^2\,\rho
&\!\!=\!\!&
e^{-\lambda}
\left[\left(r^2\,\nu''+r^2\,\frac{\nu'^2}{2}+2\,r\,\nu'+1\right)
-r\,\lambda'\left(r\,\frac{\nu'}{2}+1\right)
\right]
-1
\nonumber
\\
&&
-r^2\,k^2\left[
3\,p
-\frac{\rho}{\sigma}\left(\rho+3\,p\right)
\right]
\ ,
\end{eqnarray}
where the l.h.s.~would be the standard GR equation if the extra-dimensional terms
in the r.h.s.~vanished.
It is clear that not all of the latter terms are manifestly bulk contributions, 
since only the high-energy terms are explicitly proportional to $\sigma^{-1}$.
The general solution is given by
\begin{eqnarray}
\label{primsol}
e^{I(r)}\,e^{-\lambda(r)}
=
\int_{r_0}^r
\frac{2\,x\,e^{I(x)}}{x\,\nu'+4}
\left\{\frac{2}{x^2}
-k^2\left[
\rho-3\,p-\frac{1}{\sigma}
\left(\rho^2+3\,\rho\,p\right)
\right]
\right\}
dx+
\beta
\ ,
\end{eqnarray}
with 
\begin{eqnarray}
\label{I}
I(r)
\equiv
\int_{r_0}^{r}
\frac{2\,x^2\,\nu''+x^2\,{\nu'}^2+4\,x\,\nu'+4}{x\,(x\,\nu'+4)}\,
dx
\ ,
\end{eqnarray}
where $r_0$ and the integration constant $\beta$ will have to be determined according
to the specific system at hand.
For example, for a star centred around $r=0$, we will have $r_0=0$.
\par
Given a solution $\{p,\rho,\nu\}$ of the conservation equation~(\ref{con1}), 
we could determine the corresponding $\lambda$, ${\cal P}$ and ${\cal U}$ by means of
Eqs.~(\ref{primsol}), (\ref{pp}) and (\ref{uu}) respectively.
However, it was shown in Ref.~\cite{jo1} that this way does not lead in general to a metric
function having the expected form~(\ref{g11def}), which now reads
\begin{eqnarray}
\label{expectx} 
e^{-\lambda(r)}
=
1-\frac{k^2}{r}
\int_0^r
x^2\,\rho\,dx
+\frac{1}{\sigma}({\rm bulk\ effects})
\ .
\end{eqnarray}
In turn, if Eq.~(\ref{expectx}) does not hold,
the GR limit cannot be recovered by simply taking $\alpha\equiv \sigma^{-1}\to 0$.
The problem originates from the solution~(\ref{primsol}), which contains
a mix of GR terms and non-local bulk terms that makes it impossible to regain GR
from an arbitrary brane-world solution.
\section{The Minimal Geometric Deformation}
\label{s5}
\setcounter{equation}{0}
As we argued in Section~\ref{s2}, GR must be recovered in the limit 
$\alpha\equiv \sigma^{-1}\to 0$.
Since a brane-world observer should also see a geometric deformation due to the existence
of the fifth dimension, we restrict our search to $\{p,\rho,\nu\}$ that, beside being conserved according
to Eq.~\eqref{con1}, are such that the corresponding metric function $\lambda$ takes
the form~\eqref{expectx}, which we rewrite as
\begin{eqnarray}
\label{expectg}
e^{-\lambda(r)}
=
\mu(r)+f(r)
\ ,
\end{eqnarray}
where 
\begin{equation}
\label{standardGR}
\mu(r)
\equiv
1-\frac{k^2}{r}\int_0^r x^2\,\rho\, dx
=1-\frac{2\,m(r)}{r}
\end{equation}
is the standard GR solution containing the mass function $m$, and the unknown {\it geometric deformation}, 
described by the function $f$, stems from two sources: the extrinsic curvature and the 5-dimensional
Weyl curvature.
\par
Upon substituting~(\ref{expectg}) into Eq.~(\ref{edlrw}), we obtain the first order differential equation
\begin{equation}
\label{diffeqtof}
f'
+
\frac{2\,r^2\,\nu''+r^2\,{\nu'}^2+4\,r\,\nu'+4}{r\,(r\,\nu'+4)}\,f
=
\frac{2\,r}{r\,\nu'+4}
\left[
\frac{k^2}{\sigma}\,\rho\,(\rho+3\,p)-H(p,\rho,\nu)
\right]
\ ,
\end{equation}
where 
\begin{equation}
\label{H}
H(p,\rho,\nu)
\equiv
3\,k^2\,p
-\left[\mu'\left(\frac{\nu'}{2}+\frac{1}{r}\right)
+\mu\left(\nu''+\frac{\nu'^2}{2}+\frac{2\,\nu'}{r}+\frac{1}{r^2}\right)
-\frac{1}{r^2}\right]
\ .
\end{equation}
Solving Eq.~(\ref{diffeqtof}) yields
\begin{equation}
\label{fsolution}
f(r)
=
e^{-I(r)}\int_{0}^r
\frac{2\,x\,e^{I(x)}}{x\,\nu'+4}
\left[H(p,\rho,\nu)
+\frac{k^2}{\sigma}\left(\rho^2+3\,\rho\, p\right)\right]dx
+\beta(\sigma)\,e^{-I(r)}
\ ,
\end{equation}
where the function $I=I(r)$ is again given in Eq.~(\ref{I}) with $r_0=0$ and the integration
constant $\beta$ is taken to depend on the brane tension in such a way
that it vanishes in the GR limit $\sigma^{-1}\to 0$.
\par
Upon comparing with Eqs.~(\ref{g11}) and~(\ref{g22}) with $\mu$ given in
Eq.~\eqref{standardGR}, one can see that the non-local function 
\begin{equation}
\label{H2}
H(p,\rho,\nu)
=
3\,k^2 p-\left.\left(2\,G_2^2+G_1^1\right)\right|_{\sigma^{-1}\to 0}
\ ,
\end{equation}
which clearly corresponds to an anisotropic term, since it vanishes
in the GR case with a perfect fluid.
This feature can also be seen explicitly from computing Eq.~(\ref{pp}), which now reads
\begin{eqnarray}
\label{ppf}
\frac{6\,{\cal P}}{k^2}
=
-\frac{1}{r^2}
+\left(\frac{1}{r^2}+\frac{\nu'}{2\,r}-\frac{\nu''}{2}-\frac{{\nu'}^2}{4}\right)(\mu+f)
-\left(\nu'+\frac{2}{r}\right)\frac{\mu'+f'}{4}
\ .
\end{eqnarray}
\par
\begin{figure}[t]
\center
\includegraphics[scale=0.3]{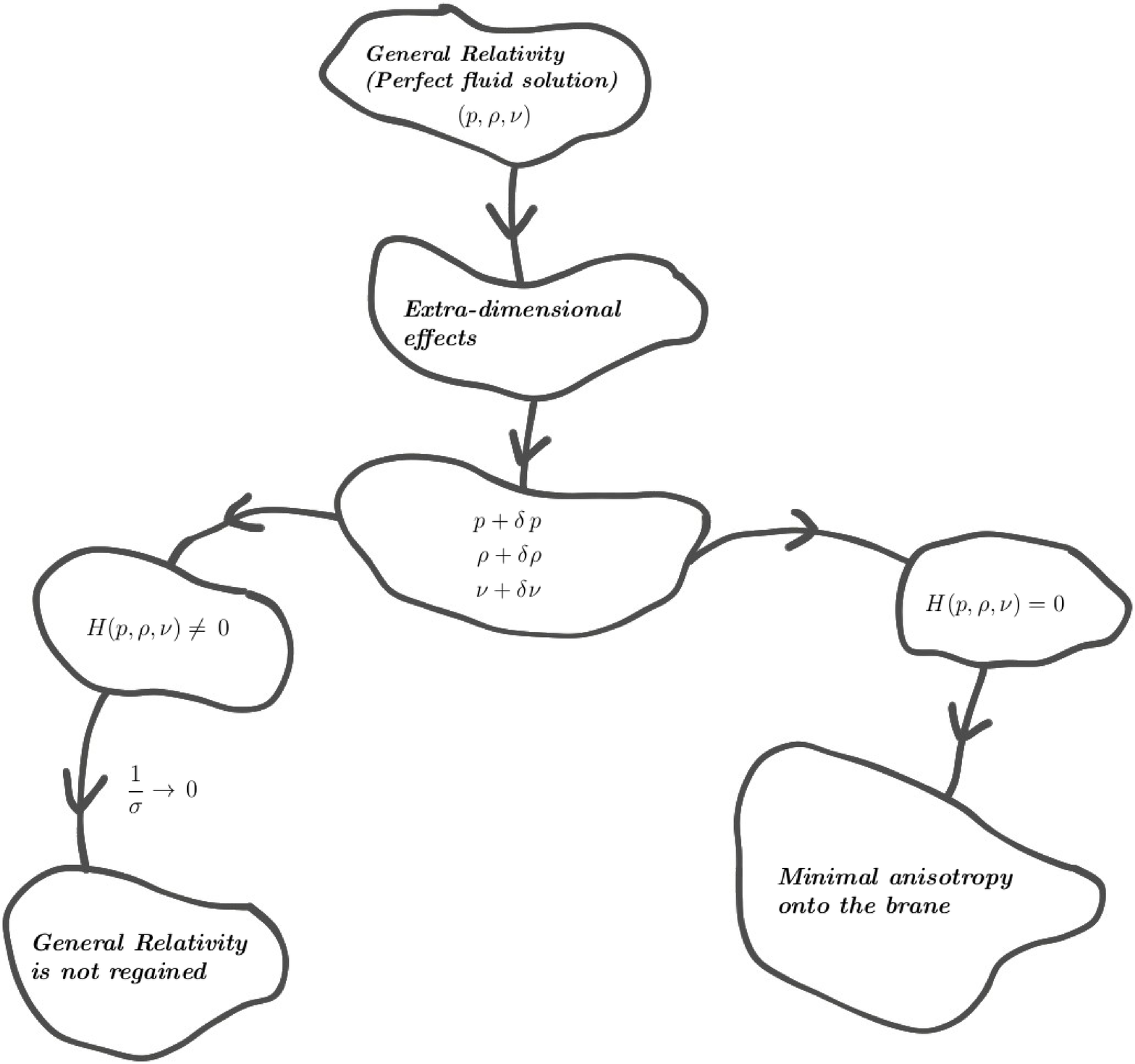}
\\
\centering\caption{For $H(p,\rho,\nu)=0$, the extra-dimensional
effects on the variables $(p,\rho,\nu)$, namely $(\delta\,p,\delta\rho,\delta\nu)$, do not produce
any further anisotropy.
Hence the anisotropy remains minimal onto the brane, and this is what ensures the low-energy
limit is given by GR. }
\label{fig1-4}      
\end{figure}

In order to recover GR, the geometric deformation~(\ref{fsolution}) must vanish for $\sigma^{-1}\to 0$.
This is achieved provided $\beta(\sigma)\to 0$ and
\begin{equation}
\label{constraintf}
\lim_{{\sigma}^{-1}\to 0}
\int_0^r\frac{2\,x\,e^{I(x)}}{x\,\nu'+4)}\,H(p,\rho,\nu)\,dx
=0\ ,
\end{equation}
which can be interpreted as a constraint for physically acceptable solution.
A crucial observation is now that, for any given (spherically symmetric) perfect fluid solution of GR,
one obtains
\begin{equation}
H(p,\rho,\nu)=0
\ ,
\label{H=0}
\end{equation}
which means that every (spherically symmetric) perfect fluid solution of GR will produce a
{\it minimal\/} deformation in the radial metric component~(\ref{expectg}) given by
\begin{equation}
\label{fsolutionmin}
f^{*}(r)
=
\frac{2\,k^2}{\sigma}\,
e^{-I(r)}\int_0^r
\frac{x\,e^{I(x)}}{x\,\nu'+4}\left(\rho^2+3\,\rho\, p\right)
dx
\ .
\end{equation}
We would like to stress that this deformation is minimal in the sense that all sources of the deformation
in Eq.~(\ref{fsolution}) have been removed, except for those produced by the density and pressure,
which will always be present in a realistic stellar distribution~\footnote{There is a MGD solution
in the case of a dust cloud, with $p=0$, but we will not consider it in the present work.}.
The function $f^{*}(r)$ will therefore produce, from the GR point of view, a ``minimal distortion'' of
the GR solution one wishes to consider.
The corresponding anisotropy induced on the brane is also minimal, as can be seen
from comparing its explicit form obtained from Eq.~(\ref{pp}),
\begin{eqnarray}
\label{ppf3}
\frac{6\,{\cal P}}{k^2\,\sigma}
=
\left(\frac{1}{r^2}+\frac{\nu'}{2\,r}-\frac{\nu''}{2}-\frac{{\nu'}^2}{4}\right)f^{*}
-\left(\nu'+\frac{2}{r}\right)\frac{(f^*)'}{4}
\ ,
\end{eqnarray}
with the general expression~(\ref{ppf}).
In particular, the constraint~(\ref{H=0}) represents a condition of isotropy in GR, 
and it therefore becomes a natural way to generalise perfect fluid solutions
(GR) in the context of the brane-world in such a way that the inevitable anisotropy
induced by the extra dimension vanishes for $\sigma^{-1}\to 0$ (see Fig.~\ref{fig1-4}).
\section{Matching condition for stellar distributions}
\label{s6}
\setcounter{equation}{0}
An important aspect regarding the study of stellar distributions is the matching conditions
at the star surface ($r=R$) between the interior ($r<R$)  and the exterior ($r>R$) geometry.
\par
In our case, the interior stellar geometry is given by the MGD metric 
\begin{equation}
ds^{2}
=
e^{\nu^{-}(r)}\,dt^{2}
-\left(1-\frac{2\,\tilde{m}(r)}{r}\right)^{-1}dr^2
-r^{2}\left(d\theta ^{2}+\sin {}^{2}\theta d\phi ^{2}\right)
\ ,
\label{mgdmetric}
\end{equation}
where the interior mass function is given by
\begin{equation}
\label{effecmass}
\tilde{m}(r)
=
m(r)-\frac{r}{2}\,f^{*}(r)
\ , 
\end{equation}
with $m$ given by the standard GR expression~(\ref{standardGR}) and
$f^{*}$ the minimal geometric deformation in Eq.~(\ref{fsolutionmin}).
Moreover, Eq.~(\ref{fsolutionmin}) implies that
\begin{equation}
f^{\ast }(r)\geq 0
\ ,
\label{f*>0}
\end{equation}
so that the effective interior mass~(\ref{effecmass}) is always reduced by the
extra-dimensional effects.
\par
The inner metric~(\ref{mgdmetric}) should now be matched with an outer
vacuum geometry, with $p^+=\rho^+=0$, but where we can in general have
a Weyl fluid described by the scalars $\mathcal{U}^{+}$ and $\mathcal{P}^{+}$~\cite{germ}.
The outer metric can be written as
\begin{equation}
ds^{2}
=
e^{\nu^{+}(r)}\,dt^{2}
-e^{\lambda^{+}(r)}\,dr^{2}
-r^{2}\left(d\theta ^{2}+\sin {}^{2}\theta d\phi ^{2}\right)
\ ,
\label{genericext}
\end{equation}
where the explicit form of the functions $\nu ^{+}$ and $\lambda ^{+}$ are
obtained by solving the effective 4-dimensional vacuum Einstein equations
\begin{equation}
R_{\mu \nu }-\frac{1}{2}\,R^\alpha_{\ \alpha}\,g_{\mu \nu}
=
\mathcal{E}_{\mu \nu }
\qquad
\Rightarrow
\qquad R^\alpha_{\ \alpha}=0
\ ,
\end{equation}
where we recall that extra-dimensional effects are contained in the
projected Weyl tensor $\mathcal{E}_{\mu \nu }$.
Only a few such analytical solutions are known to
date~\cite{MGDextended1,MGDextended2,dmpr,germ,fabbri}.
Continuity of the first fundamental form at the star surface $\Sigma$
defined by $r=R$ reads
\begin{equation}
\left[ ds^{2}\right] _{\Sigma }=0
\ ,
\label{match1}
\end{equation}
where 
$[F]_{\Sigma }\equiv F(r\rightarrow R^{+})-F(r\rightarrow R^{-})\equiv F_{R}^{+}-F_{R}^{-}$,
for any function $F=F(r)$, which yields 
\begin{equation}
{\nu ^{-}(R)}
=
{\nu ^{+}(R)}
\ ,
\label{ffgeneric1}
\end{equation}
and
\begin{equation}
1-\frac{2\,M}{R}+f_{R}^{*}
=
e^{-\lambda ^{+}(R)}
\ ,
\label{ffgeneric2}
\end{equation}
where $M=m(R)$.
Likewise, continuity of the second fundamental form at the star surface reads
\begin{equation}
\left[G_{\mu \nu }\,r^{\nu }\right]_{\Sigma }
=
0
\ ,
\label{matching1}
\end{equation}
where $r_{\mu }$ is a unit radial vector.
Using Eq.~(\ref{matching1}) and the general Einstein equations~(\ref{4Dein}),
we then find 
\begin{equation}
\left[T_{\mu \nu }^{\rm eff}\,r^{\nu }\right]_{\Sigma}
=
0
\ ,
\label{matching2}
\end{equation}
which leads to 
\begin{equation}
\left[ p
+\frac{1}{\sigma }\left( \frac{\rho ^{2}}{2}+\rho \,p+\frac{2}{k^{4}}\,\mathcal{U}\right)
+\frac{4\,\mathcal{P}}{k^4\,\sigma }\right]_{\Sigma }
=
0
\ .
\label{matching3}
\end{equation}
Since we assumed the star is only surrounded by a Weyl fluid characterised
by $\mathcal{U}^{+}$, $\mathcal{P}^{+}$, this matching condition takes the final form 
\begin{equation}
p_{R}
+\frac{1}{\sigma }\left( \frac{\rho _{R}^{2}}{2}+\rho _{R}\,p_{R}+\frac{2}{k^{4}}\,\mathcal{U}_{R}^{-}\right)
+\frac{4\,\mathcal{P}_{R}^{-}}{k^4\,\sigma }
=
\frac{2\,\mathcal{U}_{R}^{+}}{k^4\,\sigma }
+\frac{4\,\mathcal{P}_{R}^{+}}{k^4\,\sigma }
\ ,
\label{matchingf}
\end{equation}
where $p_{R}\equiv p^{-}(R)$ and $\rho _{R}\equiv \rho^{-}(R)$.
Finally, by using Eqs.~(\ref{uu}) and (\ref{ppf3}) in the condition~(\ref{matchingf}),
the second fundamental form can be written in terms of the MGD at the
star surface, denoted by $f_{R}^{\ast }$, as 
\begin{equation}
p_{R}
+\frac{f_{R}^{*}}{k^2}\left( \frac{\nu _{R}^{\prime }}{R}+\frac{1}{R^{2}}\right)
=
\frac{2\,\mathcal{U}_{R}^{+}}{k^4\,\sigma }
+\frac{4\,\mathcal{P}_{R}^{+}}{k^4\,\sigma }
\ ,
\label{sfgeneric}
\end{equation}
where $\nu _{R}^{\prime }\equiv \partial _{r}\nu^{-}|_{r=R}$. 
Eqs.~(\ref{ffgeneric1}), (\ref{ffgeneric2}) and~(\ref{sfgeneric}) are the necessary
and sufficient conditions for the matching of the interior MGD metric~(\ref{mgdmetric})
to a spherically symmetric ``vacuum'' filled by a brane-world Weyl fluid.
\par
The matching condition~(\ref{sfgeneric}) yields an important result:
if the outer geometry is given by the Schwarzschild metric, one must have
$\mathcal{U}^{+}=\mathcal{P}^{+}=0$, which then leads to 
\begin{equation}
p_{R}
=
-\frac{f_{R}^{\ast }}{k^2}
\left( \frac{\nu _{R}^{\prime }}{R}+\frac{1}{R^{2}}\right)
\ .
\label{pnegative}
\end{equation}
Given the positivity of $f^*$, Eq.~(\ref{f*>0}), an outer Schwarzschild
vacuum can only be supported in the brane-world by exotic stellar
matter, with $p_{R}<0$ at the star surface.
\section{The recipe}
\label{s7}
\setcounter{equation}{0}
Let us conclude this brief introduction of the MGD approach by listing the basic steps
to implement it:
\begin{description}
\item[Step 1:] 
pick a known perfect fluid solution $\{p,\rho,\nu\}$ of the conservation equation~(\ref{con1}).
This solution will ensure that $H(p,\rho,\nu)=0$ and the radial metric component $\lambda$ will be
given by Eq.~(\ref{expectg}) with $f=f^*$ in Eq.~(\ref{fsolutionmin}). 
The GR solution will be recovered in the limit $\sigma^{-1}\to 0$ by construction.
\item[Step 2:]
determine the Weyl functions ${\cal P}$ and ${\cal U}$ from Eqs.~(\ref{pp}) and~(\ref{uu}).
\item[Step 3:]
use the second fundamental form given in Eq. (\ref{sfgeneric}) to express any GR constant $C$ as a function of the brane tension $\sigma$, that is, $C(\sigma)$. Then we are able to find the bulk effect on pressure $p$ and density $\rho$, that is, $p(\sigma)$ and $\rho(\sigma)$.
%
%
%
\end{description}
\section{Conclusions}
\label{s8}
\setcounter{equation}{0}
In the context of the Randall-Sundrum brane-world, a brief and detailed description of the
basic elements of the MGD was presented.
The explicit form of the anisotropic stress ${\cal P}$ was obtained in terms of the
geometric deformation $f$ undergone by the radial metric component,
thus showing the role played by this deformation as a source of anisotropy
inside stellar distributions.
It was shown that this geometric deformation is minimal when a GR solution is considered,
therefore any perfect fluid solution in GR belongs to a subset of brane-world solutions
producing a minimal anisotropy onto the brane.
It was shown that with this approach it is possible to generate the brane-world version of
any known GR solution, thus overcoming the non-closure problem of the effective
4-dimensional Einstein equations.
A simple recipe showing the basic steps to implement the MGD approach was finally presented. A final natural question arises: is the MGD an useful approach to deal only with the effective 4-dimensional Einstein equations in the brane-world context? The answer is no. Indeed, we have found \cite{conf} that any modification of general relativity can be studied
by the MGD provided that such modification can be represented by a traceless energy-momentum
tensor. This mean that the MGD is particularly useful as long as the new gravitational sector is  associated with a conformal gravitational sector.  
\section*{Competing Interests}
\par
The authors declares that there is no conflict of interest regarding the publication of this paper.
\section{Acknowledgements}
A.S. is partially supported by Project Fondecyt 1161192, Chile.

\end{document}